\documentstyle[11pt,epsfig]{article}
\begin{document}
\begin{center}
{\bf Photometric Observation and Period Study of GO Cygni}\\ \ \\
{\small {\bf S.M. ZABIHINPOOR}\\
{\it Biruni Observatory, Shiraz University, Shiraz 71454, Iran}\\ \ \\

{\bf A. DARIUSH}\\
{\it Institute for Advanced Studies in Basic Sciences (IASBS), Zanjan, Iran; \\E-mail: dariush@iucaa.ernet.in}\\ \ \\

{\bf N. RIAZI}\\
{\it Physics Department and Biruni Observatory, Shiraz University, Shiraz 71454, Iran; \\E-mail: riazi@physics.susc.ac.ir }}\\ \ \\

\end{center}


\begin{abstract}
Photometric observations of GO Cyg were performed during the
July-October 2002, in B and V bands of Johnson system. Based on
Wilson's model, the light curve analysis were carried out to find
the photometric elements of the system.  The O-C diagram which is
based on new observed times of minima  suggests a negative rate of
period variation ($\frac{dP}{dt}<0$) for the system.
\end{abstract}



\section{Introduction}
GO Cygni (HD196628; $\alpha:20^h37^m$,$\delta:+35^\circ 26'$) has
been classified as a $\beta$-Lyr type eclipsing binary of short
period of about $P=0^d.71$ whose visual magnitude varies from a
maximum light value of $V_{max}=+8.6$ to a minimum light value of
$V_{min}=+9.25$ and has a period variation. The results of the
light curve analysis and the period study can  be found in many
papers such as Ovenden (1954), Popper (1957), Rovithis et
al.(1990), Hall \& Louth (1990), Sezer et al.(1993), Edalati \&
Atighi (1997), Rovithis et al.(1997), and Oh \& Ra (1998).

The structure of the paper is as follows. In Sec.2 we introduce
the process of observations and find the times of minima. In Sec.3
the period study of the system is presented. The Wilson-Devinney
model has been used to get the orbital and physical parameters of
the system which are given in Sec.4. In Sec.5, the absolute
dimensions of the system is calculated by the use of spectroscopic
data together with the parameters estimated from the present
study. A brief discussion and our main conclusions are presented
in Sec.6.


\section{{Observations and times of minima}}
We present the  observed light-curve of GO Cygni, together with
the new times of minima from the photometric observations carried
out at Biruni Observatory. The 20 inches cassegrainian reflector
which is equipped with an uncooled RCA4509 multiplier phototube,
was used for differential photometry in B($\lambda=4200$\AA) and
V($\lambda=5500$\AA) filters of Johnson's system. The observations
were performed during Aug./Sep. 2002 versus HD 197292 and HD
197346 as the comparison and the check stars. In each observing
run in B or V filter, the integration time was fixed for 10
seconds. The observed data were transferred to computer using an
A/D converter. The reduction of data and atmospheric corrections
were done using REDWIP code developed by G. P. McCook to obtain
the complete light-curves in the two filters (Fig. 1). The times
of minima in {\it Heliocenteric Julian Date} (two primary and two
secondary) were calculated by fitting a Lorentzian function to the
observed minima data points (Dariush et al. 2003). Fig. \ref{fit},
shows a sample Lorentzian fit to the observed primary minimum of
GO Cyg on Aug. 31 in filter B.
\begin{figure}
\label{lc}
\includegraphics[scale=0.45,angle=-90]{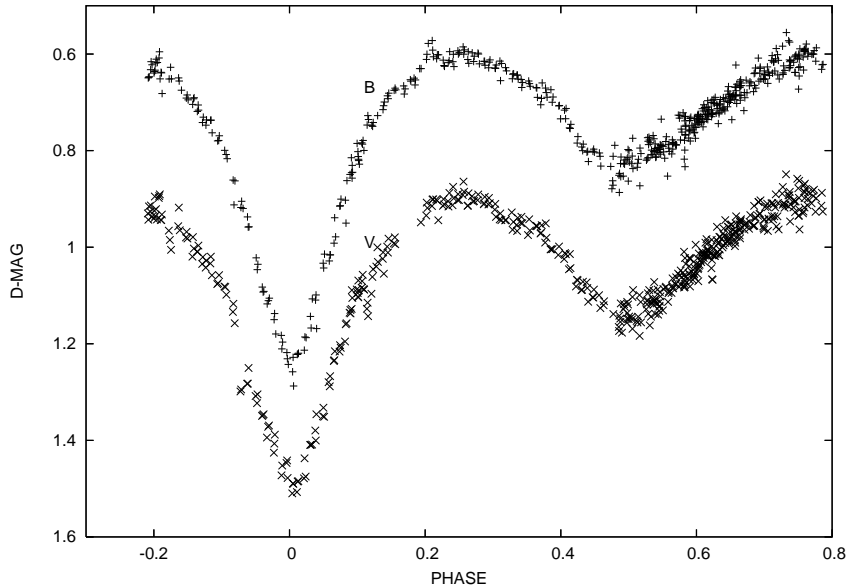}
\caption{Observed light curve of GO Cyg in the B and V filters.}
\end{figure}
\begin{figure}
\label{fit}
\includegraphics[scale=0.45,angle=-90]{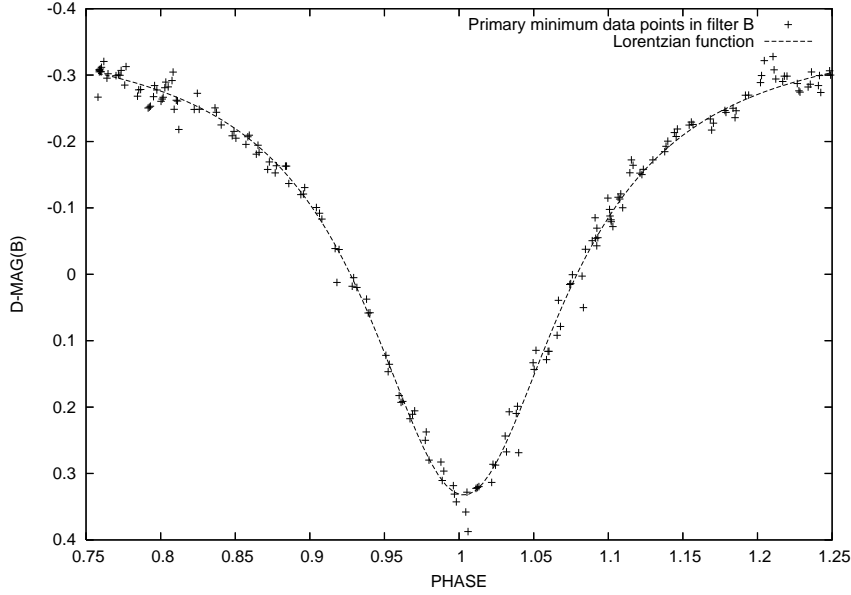}
\caption{A sample Lorentzian fit to the observed primary minimum
data on Aug. 31 in filter B.}
\end{figure}


\section{Period study}
\begin{figure}
\label{oc}
\includegraphics[scale=1.5]{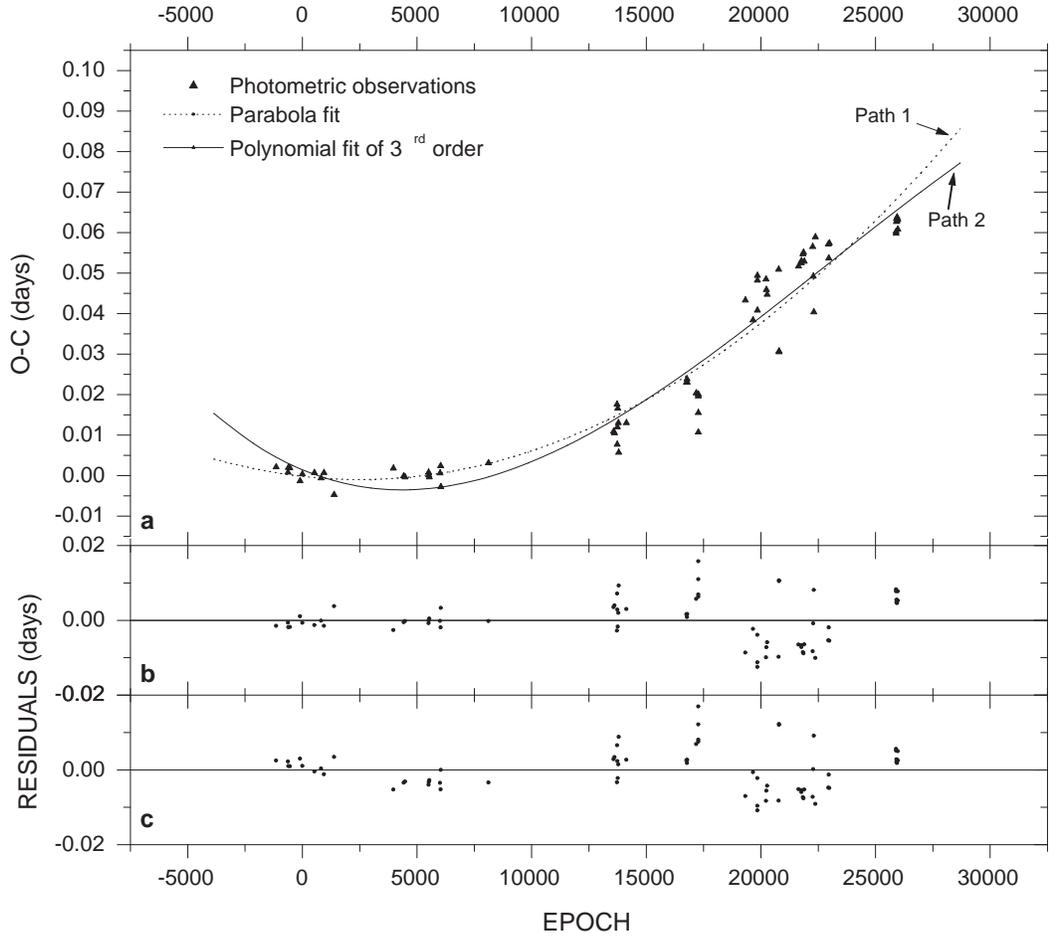}
\caption{\textbf{(a)} Plot of (O-C)'s versus epoch for all the
observed photoelectric times of primary and secondary minima from
1951 to 2002. The residuals are calculated according to equation
\ref{ephem1}. The dotted line and solid line represent a 3$^{rd}$
order polynomial ($j$=3 in equation \ref{poly}) and a parabola
($j$=2 in equation \ref{poly}) fitting functions  to all of the
residuals respectively (see Table 1). \textbf{(b)} Residuals
between the observed O-C differences and the best fitted
parabola.\textbf{(c)} Residuals between the observed O-C
differences and the best fitted 3$^{rd}$ polynomial. }
\end{figure}

Though a constant period for GO Cyg was first measured by
Szczyrbak (1932) other studies such as Purgathofer \& Prochazka
(1967), Cester et al. (1979), Sezer et al. (1985), Jones et al.
(1994),  Rovithis et al.(1997), and Edalati \& Atighi (1997)
suggested a period variation in GO Cyg. For all the observed
photoelectric times of primary and secondary minima from 1951 to
2002 the computed (O-C)'s versus epoch is plotted in Fig. 3. For
each minima, the residuals are calculated according to the
ephemeris given by Hall \& Louth (1990)
\begin{equation}
\label{ephem1}
 Min.I=HJD 2433930.40561+0^{d}.71776382 \times E.
\end{equation}
In equation \ref{ephem1}, $E$ is the number of cycles. Since the
period is changing with time an extra {\it quadratic term} can be
added to equation \ref{ephem1} and rewrite it as
\begin{equation}
\label{ephem2}
 Min.I=HJD 2433930.40561+0^{d}.71776382 \times E+\xi \times E^2.
\end{equation}
From the method of least square the coefficient of quadratic term
in equation \ref{ephem2} is easily found to be $\xi=9.35\times
10^{-11}$. As it is seen from Fig. 3a, the O-C residuals suggest a
polynomial function in the form of
\begin{equation} \label{poly}
\Delta T(E)=\sum_{j=1}^n c_jE^j,
\end{equation}
where $\Delta T(E)$ is the differences between the observed and
calculated times of minima for any cycle $E$. In the case of GO
Cyg, the 3$^{rd}$ order polynomials ($j=3$ in equation \ref{poly})
seems to be fitted better than the parabola ($j=2$ in equation
\ref{poly}) to the observed times of minima. The value of the
polynomial coefficients $c_j$ and the {\it rms error} of the
approximation are listed in Table 1. The residuals of the observed
O-C from $\Delta T(E)$ which do not deviate more than $\pm 0.01$d
are shown in Fig. \ref{oc}b and \ref{oc}c. Using equation
\ref{poly}, the real period of the GO Cyg  and its relative rate
of change can be calculate according to the following formula (see
Kalimeris et al. 1994)
\begin{equation}
\label{per}
P(E)=P_e+\Delta T(E) - \Delta T(E-1),
\end{equation}
\begin{equation}
\label{dper}
\dot{P}(E)=\frac{dP}{dE}=\{\sum_{j=0}^{n-1}(j+1)c_{j+1}E^j-\sum_{j=0}^{n-1}(j+1)c_{j+1}(E-1)^j\},
\end{equation}
where $P_e$ is an ephemeris period given in equation \ref{ephem1}.
Fig. 4 shows the period of GO Cyg as a function of time and its
relative rate of change according to the 3$^{rd}$ order polynomial
fitting function on the O-C diagram. From Fig. 4, it is obvious
that at present time, the period is decreasing. We refer the
reader to Sec. 6 where we presented a completed discussion about
the period variations.

\begin{table}
\label{occ}
 \begin{tabular}{lcc}\hline
 Coefficient&  Second                   &  Third                    \\\hline
 $c_0$      &  $-2.92771\times10^{-4}$  &  $1.45\times10^{-3}$      \\
 $c_1$      &  $-6.38377\times10^{-7}$  &  $-2.3711\times10^{-6}$   \\
 $c_2$      &  $1.26577\times10^{-10}$  &  $3.01298\times10^{-10}$  \\
 $c_3$      &                           &  $-4.41579\times10^{-15}$ \\
 rms error  &     0.00625               &  0.00597            \\
 \hline
\end{tabular}
\caption[]{Coefficients of second and third order approximation of
the (O-C) diagram of GO Cyg.}
\end{table}

\begin{figure}
\label{period}
\includegraphics[scale=1.3]{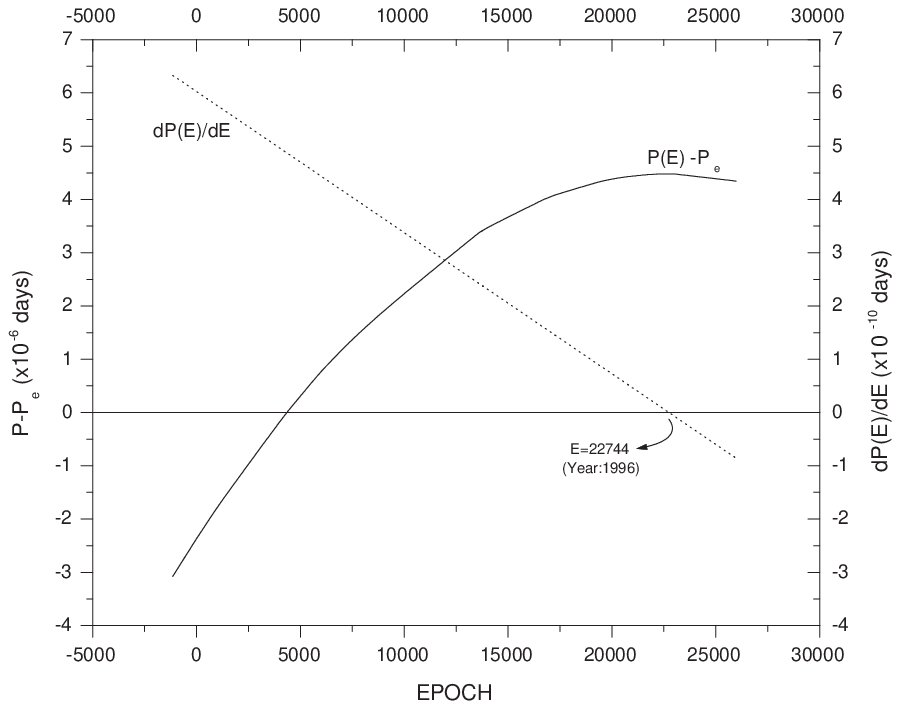}
\caption{The period of GO Cyg as a function of time (continues
line) and its relative rate of change (dotted line) according to
the equations \ref{per} and \ref{dper} with $j$=3. The difference
$P-P_e$ is referred to the ephemeris period $P_e$=0.71776382d
which is given in equation \ref{ephem1}.}
\end{figure}


\section{{Light curve analysis}}
The photometric solution and light curve analysis is done using
the Wilson and Devinney's (1971) model which is based on Roche
model. Before doing analysis we binned the observed data in B and
V filters presented in Fig. 1 in equal phase intervals and applied
Wilson's {\bf LC} code in {\bf mode 4}. This mode is for
semi­detached binaries with star 1 accurately filling its limiting
lobe, which is the classical Roche lobe for synchronous rotation
and a circular orbit, but is different from the Roche lobe for
non­synchronous and eccentric cases. The applied constraints are
that $\Omega_1$ has the lobe filling value and that $L_2$ is
coupled to the temperatures. The {\bf LC} code consists of a main
FORTRAN program for generating light curves. Solutions were
obtained for each of the B and V filters, separately. The
preliminary values for the orbital and physical elements of the
system were adopted from the previous studies. The third light and
eccentricity were supposed to be equal to zero ($e=0, l_3=0$). In
order to optimize the parameters, an  auxiliary computer programme
was written to compute the sum of squares of the residuals (SSR)
in both B and V filters, using {\bf LC} outputs.   The set of
parameters ($i,q,T_2,\Omega_1,\Omega_2,L_1,$ and $L_2$) given in
 Table 2, are derived after optimization in the way as we
 described. Fig. 5 and 6 show the
 optimized theoretical light curves
together with the observed light curves in B and V filters. These
theoretical curves, correspond to the optimized parameters given
in Table 2. In Table 4 we tabulated our results (as an average in
B and V filters) together with previous photometric solutions from
other sources.

\begin{figure}[h]
\label{fitb}
\includegraphics[scale=0.45,angle=-90]{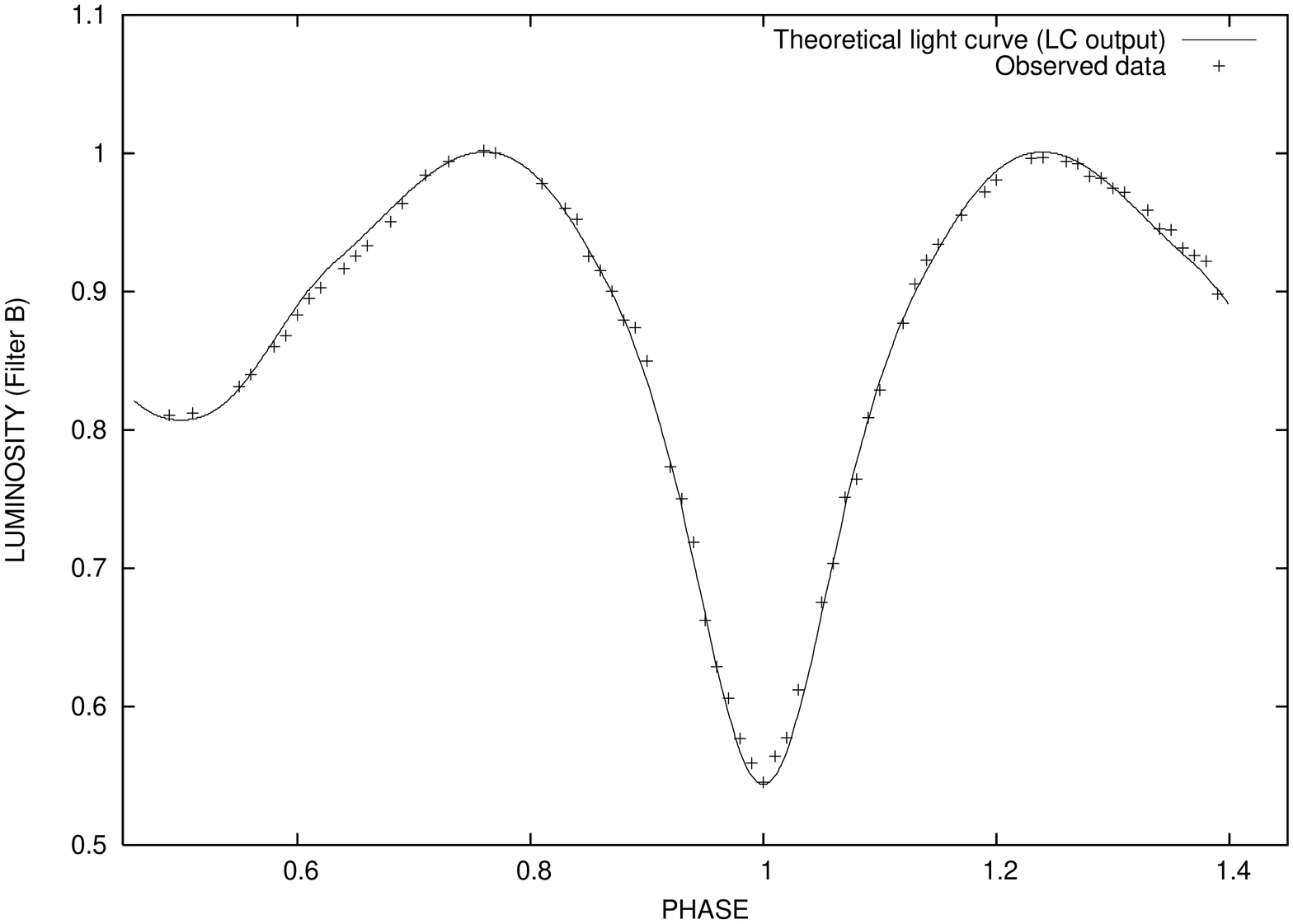}
\caption{Theoretical light curve ({\bf LC} output; solid line) in
filter B according to the parameters given  in Table 2. Points are
the observational data.}
\end{figure}
\begin{figure}[h]
\label{fitv}
\includegraphics[scale=0.45,angle=-90]{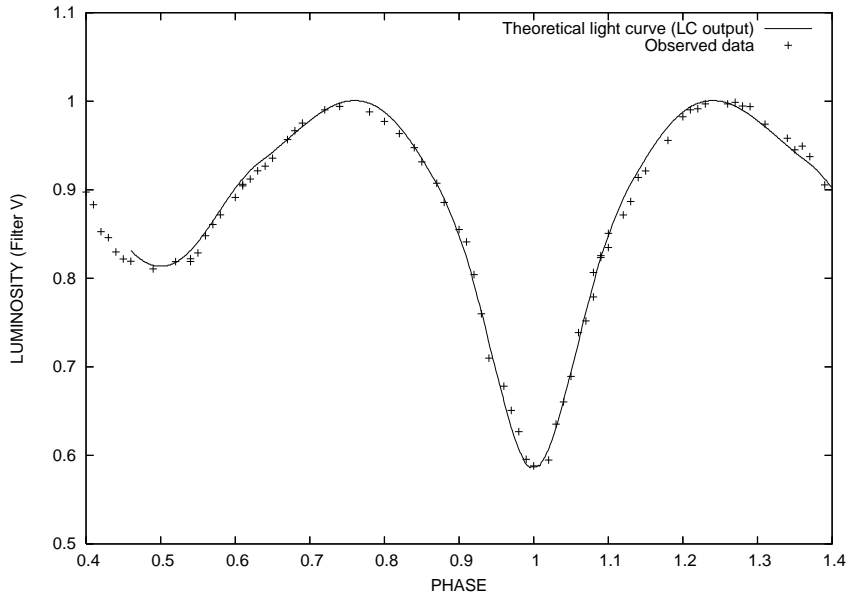}
\caption{Same as Fig. 5 but in V filter}
\end{figure}

\begin{table}\label{Tab1}
 \begin{tabular}{lll}\hline
 Parameter           &  Filter B               &  Filter V \\\hline
 $\lambda$(\AA)      &  4200                   &  5500     \\
 $i$(degree)         &  77.72$\pm 0.02$        &  78.38$\pm 0.02$    \\
 $q(\frac{M_2}{M_1})$&  0.428$\pm 0.001$       &  0.428$\pm 0.001$    \\
 $\Omega_1$          &  2.734                  &  2.741    \\
 $\Omega_2$          &  2.905$\pm 0.005$       &  2.902$\pm 0.002$    \\
 $T_1$(fixed)        &  10350                  &  10350    \\
 $T_2$               &  6528$\pm 30$           &  6404$\pm 40$  \\
 $A_1$               &  1                      &  1        \\
 $A_2$               &  0.5                    &  0.5      \\
 $g_1$               &  1                      &  1        \\
 $g_2$               &  0.32                   &  0.32     \\
 $x_1$               &  0.82                   &  0.58     \\
 $x_2$               &  0.79                   &  0.51     \\
$\frac{L_1}{L_1+L_2}$&  0.965                  &  0.947    \\
$\frac{L_2}{L_1+L_2}$&  0.035                  &  0.053    \\
 $r_1$(pole)         &  0.427                  &  0.427    \\
 $r_1$(point)        &  0.586                  &  0.586    \\
 $r_1$(side)         &  0.455                  &  0.455    \\
 $r_1$(back)         &  0.482                  &  0.482    \\
 $r_2$(pole)         &  0.259                  &  0.259    \\
 $r_2$(point)        &  0.300                  &  0.300    \\
 $r_2$(side)         &  0.267                  &  0.267    \\
 $r_2$(back)         &  0.285                  &  0.285    \\
 $\Sigma(O-C)^2$     &  0.00172                &  0.00214  \\
\hline
\end{tabular}
\caption[]{Optimized parameters of GO Cygni.}
\end{table}


\section{{Absolute dimensions}}
By combining the photometric and spectroscopic data, it is
possible to estimate the absolute physical and orbital parameters
of the system. The first spectroscopic elements of the system were
prepared by Pearce (1933) and a more recent spectroscopic orbit
was obtained in 1988 by Holmgren using the cross-correlation
techniques ( Sezer et al., 1992). We use the spectroscopic
elements of Holmgren together with the orbital parameters and
physical parameters which have been calculated already from our
observation to calculate the absolute dimensions of GO Cyg. The
results of these calculations based on equations \ref{param} (see
Kallrath \& Milone, 1999) is given in Table 3. In equations
\ref{param}, $f(M_1,M_2,i)$ is mass function, $M_{tot}=M_1+M_2$ is
the total mass, $K_1$ is the radial velocity amplitude of
component 1, $a$ is the relative semi-major axis, and $g$ is the
acceleration due to gravity at the surface.

\begin{equation}
\label{param} \  \left\{
\begin{array}{lllllll}
M_1=\frac{f(M_1,M_2,i)(1+q)^2}{q^3\sin^3i}  \\

f(M_1,M_2,i)=\frac{P}{2\pi G}(1-e^2)^{\frac{3}{2}}{K_1}^3\\

\frac{M_2}{M_{\odot}}=q(\frac{M_1}{M_{\odot}})\\

g=G\frac{M}{R^2}=G\frac{M/M_{\odot}}{(R/R_{\odot})^2}; g_{\odot}=2.74\times10^2 \frac{m}{s^2}\\

\frac{L}{L_{\odot}}=(\frac{R}{R_{\odot}})^2(\frac{T}{T_{\odot}})^4\\

P^2=\frac{4\pi^2}{G}\frac{ a^3}{M_1+M_2}\\

a=a_1\frac{1+q}{q}=(1+q)a_2\\

\end{array} \right.
\end{equation}

\begin{table}\label{Tab2}
 \begin{tabular}{lll}\hline
 Parameter           &  Filter B   &  Filter V \\\hline
 $\lambda$(\AA)      &  4200       &  5500     \\
 $M_1(M_{\odot})$    &  2.74       &  2.82     \\
 $M_2(M_{\odot})$    &  1.73       &  1.75     \\
 $R_1(R_{\odot})$    &  2.16       &  2.17     \\
 $R_2(R_{\odot})$    &  1.09       &  1.08     \\
 $L_1(L_{\odot})$    &  58.72      &  57.54    \\
 $L_2(L_{\odot})$    &  2.17       &  2.13     \\
 \hline
\end{tabular}
\caption[]{Absolute elements of GO Cygni.}
\end{table}


\section{{Results and discussion}}
A closer look at Fig. 5 and 6, shows that the primary minimum is
symmetric around phase 1 and the secondary minimum occurs at phase
0.5. Also, the out of eclipse portion of the light curve in two
figures is the same which is consistent with the results of Sezer
et al. (1985) and Jassur \& Puladi (1995). But there is a lack of
symmetry in the secondary minimum of both figures around phase
interval of 0.6 to 0.7 which was also reported earlier (Sezer et
al., 1993 and Edalati \& Atighi, 1997). Since the O-C diagram of
GO Cyg clearly shows a variation in period ,the asymmetry in
secondary minimum is most probably due to gas streaming or mass
transfer. The result of our study support this idea that the
primary component $M_1$ has filled its Roche lobe, and the second
component ($M_2$; less massive and cooler one) is somewhat smaller
than its Roche lobe. Let the total mass of the binary system be
$M_{tot}=M_1+M_2$ and take the total angular momentum of the
orbiting pair as $J_{orb}$. In reality, the total angular momentum
$J_{tot}$ is composed of two terms $J_{orb}$ and $J_{spin}$(spin
or rotational angular momentum) such that
$J_{tot}=J_{orb}+J_{spin}$ but in practice the contribution of the
total spin angular momentum is no more than 1-2\% of $J_{orb}$, so
it can be ignored (Hilditch, 2001). Now consider a conservative
mass transfer between the two components as well as the
conservation of total orbital angular momentum $J_{orb}$.
Therefore $\dot{M}=\frac{d{M_{tot}}}{dt}=0$ or $dM_1=-dM_2$ (since
all the mass lost by one component is gained by its companion),
also $\frac{d{J_{orb}}}{dt}=0$. Using the relation between
$M_{tot}$ and $J_{orb}$ together with Kepler's third law, lead us
to the following relation (Hilditch, 2001)
\begin{equation}
\label{mdot}
\dot{M_1}=\frac{\dot{P}M_1M_2}{3P(M_1-M_2)},
\end{equation}
in which $P$ and $\dot{P}$ are -in order- the orbital period and
the rate at which the orbital period is changing and $M_1$ is mass
of the massive (and hotter) component.

Now suppose that the system follows {\it path 1} on the O-C
diagram which is defined by the parabolic fit. In this case
according to the equations \ref{per} and \ref{dper}, $\dot{P}>0$
(also $M_1>M_2$), so it is clear from equation \ref{mdot} that
$\dot{M}_1>0$ and the mass of the more massive component ($M_1$)
must increases with time. But from the Roche model we expect
$\dot{M}_1<0$ since the primary component has filled its Roche
lobe and is loosing its mass to its cooler companion $M_2$. As a
result if we consider a simple conservative model for mass
transfer with the spin angular momentum ignored then there is an
inconsistency between the rate of period change and conservative
model of mass transfer of the system. This deviation may be due to
the role of stellar winds or spin-orbital angular momentum
interchange.

On the other hand if the system follows  {\it path 2} on the O-C
diagram where is defined by a 3$^{rd}$ order polynomial, then it
is clear from Fig. 4 that at the present time and after 1996,
$\dot{P}<0$, which means that equation \ref{mdot} is valid and the
mass of the primary component is transferring to the secondary one
($\dot{M}_1<0$ and $\dot{M}_2>0$) which is in agreement with the
Roche model we considered for the GO Cyg. The point is that the
ascending branches of both parabolic and 3$^{rd}$ order polynomial
fit to the O-C data points in Fig. 3a intersect around period
cycle $E$=24000 (see also Rovithis et al. 1997), so the
observations after this time is very important to give a clear
understanding of the period change of GO Cyg and its behaviour.
The results of our observations strongly support this idea that
the rate of period change is negative but we must extend our
observational baseline in order to collect more minima to study
the behaviour of the GO Cyg O-C diagram and to see which model is
most probable. If the new times of minima still follow {\it path
2} in O-C diagram then there is no doubt that the increase in rate
of change of the period of the system has been stopped and the
system is now undergoing a decrease of period.

\begin{table}\label{tab3}
 \begin{tabular}{llllllll}\hline
Parameters & Ovenden & Mannino & Popper & Sezer et al. & Rovithis et al. & Edalati et al.& Present\\
 & (1954) &   (1963) &   (1980) &  (1993) &        (1997)$^1$  &   (1997) & study$^2$\\ \hline
 $i$(degree)           & 78.777 & 79.880 & 77.606 & 79.722 & 78.414 & 82.311 & 78.05 \\
 $T_1$                 & 10350  & 10350  & 10350  & 10350  & 10350  & 10350  & 10350\\
 $T_2$                 & 5605   & 5904   & 5966   & 6043   &  6912  & 6428   & 6452\\
 $\frac{L_1}{L_1+L_2}$ & 0.969  & 0.961  & 0.972  & 0.954  & 0.931  & 0.931  & 0.956\\
 $\frac{L_2}{L_1+L_2}$ & 0.031  & 0.039  & 0.028  & 0.034  & 0.069  & 0.068  & 0.044\\
 $r_1$(back)           & 0.482  & 0.482  & 0.482  & 0.482  & 0.555  & 0.481  & 0.482\\
 $r_1$(side)           & 0.455  & 0.455  & 0.455  & 0.455  & 0.504  & 0.455  & 0.455\\
 $r_1$(pole)           & 0.427  & 0.427  & 0.427  & 0.427  & 0.445  & 0.426  & 0.427\\
 $r_2$(back)           & 0.282  & 0.271  & 0.221  & 0.278  & 0.303  & 0.275  & 0.285\\
 $r_2$(side)           & 0.265  & 0.256  & 0.214  & 0.261  & 0.280  & 0.258  & 0.267\\
 $r_2$(pole)           & 0.257  & 0.249  & 0.210  & 0.254  & 0.265  & 0.251  & 0.259\\ \hline
\end{tabular}
\caption[]{Photometric solutions from other light curves.} {\it 1. Based on the observations in Bucharest.}\\
{\it 2. Average on B and V filters.}\\

\end{table}


\subsection*{Acknowledgements}
We gratefully acknowledge the valuable remarks
and comments of the referee. We also thanks Mr. Moein Mosleh and
Mr. Hadi Rahmani for their help with some of the observations and
calculations. This research has made use of {\bf NASA}'s
Astrophysics Data System Abstract Service and {\bf SIMBAD} data
base operated at CDS (Strasbourg, France).


\end{document}